\documentclass[12pt]{iopart}
\expandafter\let\csname equation*\endcsname\relax
\expandafter\let\csname endequation*\endcsname\relax
\usepackage[utf8]{inputenc}
\DeclareUnicodeCharacter{2212}{\textminus}
\usepackage{xcolor}
\usepackage{ulem}
\usepackage{graphicx}
\usepackage{siunitx}
\usepackage[version=4]{mhchem}
\usepackage{hyperref} 
\usepackage[capitalize]{cleveref}
\usepackage[square,numbers]{natbib}
\usepackage{iopams}
\usepackage{caption}
\usepackage{subcaption}
\usepackage{multirow}
\usepackage{booktabs}
\usepackage{makecell}
\usepackage[title]{appendix}

\begin{document}

\title[Fast and Accurate Prediction of Lattice Thermal Conductivity via ML Surrogates ]{Fast and Accurate Prediction of Lattice Thermal Conductivity via Machine Learning Surrogates}

\author{Zeyu Wang$^{1\dagger}$, Shuya Yamazaki$^{2\dagger}$, Martin Hoffmann Petersen$^{2,3\dagger}$, Masato Ohnishi$^{4,5}$, Tomiya Yamamoto$^6$, Wei Nong$^2$, Jianghai Wang$^2$, Ruiming Zhu$^3$, Masatoshi Hanai$^7$, Michimasa Morita$^2$, Toyotaro Suzumura$^7$, Zekun Ren$^3$*, Junichiro Shiomi$^{1,4,5,8}$*, and Kedar Hippalgaonkar$^{2,3}$*}

\address{1 Department of Mechanical Engineering, The University of Tokyo, Tokyo 113-8654, Japan}
\address{2 School of Materials Science and Engineering, Nanyang Technological University, Singapore 639798}
\address{3 Berkeley Education Alliance for Research in Singapore, CREATE Tower, Singapore 138602 }
\address{4 Institute of Engineering Innovation, The University of Tokyo, Tokyo 113-0032, Japan}
\address{5 The Institute of Statistical Mathematics, Research Organization of Information and Systems, Tachikawa, Tokyo 190-8562, Japan}
\address{6 Division of Materials Science, Nara Institute of Science and Technology, Ikoma, Nara 630-0192, Japan}
\address{7 Information Technology Center, The University of Tokyo, Tokyo 113-0032, Japan}
\address{8 RIKEN Center for Advanced Intelligence Project, Tokyo 103-0027, Japan}

\address{$\dagger$ Equal contribution}

\textbf{}
\ead{danny.ren@bears-berkeley.sg, shiomi@photon.t.u-tokyo.ac.jp, kedar@ntu.edu.sg}

\begin{abstract}
The appearance of generative models has opened vast chemical spaces in the design of functional materials. Although machine learning interatomic potentials (MLIPs) have substantially accelerated phonon calculations, high-fidelity prediction of lattice thermal conductivity $\kappa_{lat}$ still requires accurate treatment of anharmonic interactions, which remains a key challenge for existing potentials across novel chemical spaces. 
To address this challenge, we present a comprehensive benchmark of 15 surrogate models for predicting $\kappa_{lat}$ using the Phonix database\cite{Ohnishi2026}, which contains 6,966 entries with anharmonic phonon properties derived from first-principles calculations. Firstly, We categorize these surrogate models into three distinct groups: Physical-informed feature descriptors combined with ML models, end-to-end deep neural networks, and pre-trained MLIP-embeddings combined with ML models. By evaluating model performance across random, space-group disjoint (testing generalization to unseen crystal symmetries), and Out-Of-Distribution splits (OOD dataset that testing extrapolation to property regimes beyond the training range) based on $\kappa_{lat}$, we probe both interpolation and exploration capabilities. Our results reveal that MLIP-embedded models excel in interpolation within well-sampled regions, deep neural network models especially ALiEGNN demonstrate superior robustness in OOD regimes critical for discovering novel low-$\kappa_{lat}$. Additionally, we find a systematic degradation in performance when the structural representation is reduced. Although surrogate models exhibit lower accuracy than direct simulations using first-principles calculation, they reduce computational costs by orders of magnitude, enabling efficient high-throughput screening of thermoelectric materials with minimal loss in generative design workflows.
\end{abstract}

\section{Introduction}
In recent years, Machine Learning (ML) has emerged as a powerful tool in computational materials science\cite{Schleder2019,Butler2018,oganov2019structure,Low_TC_prediction,prediction_phonon_density,E3NN_phonon,VGNN_phonon_prediction}. In particular, Machine Learning Interatomic Potentials (MLIPs) have enabled near–first-principles accuracy at significantly reduced computational cost\cite{yang2024mattersim,batatia2023foundationmace,deng2023chgnet,unke2021machine}, while advances in generative models have enabled the exploration and design of materials in large and complex chemical spaces\cite{mattergen,diffcsp_plus,kazeev2024wyckofftransformer, cgcnn_TC}. The introduction of generative models has initiated a new paradigm: rather than relying on chemically intuitive, combinatorial searches for promising functional materials, we can now train generative models on descriptive databases and sample candidate materials directly from the learned distributions\cite{Merchant2023}. 
Although these models can be conditioned on specific chemical spaces or targeted properties, ensuring that the generated structures faithfully satisfy the desired properties typically requires extensive post hoc evaluation.

A key example of functional materials of interest is thermoelectrics, which can convert waste heat directly into electricity or operate as solid-state Peltier coolers. Despite their technological potential, thermoelectric materials and devices still have limited practical adoption due to challenges related to efficiency, material cost, and large-scale manufacturability\cite{Shi2025,Wang2025_progress}. The efficiency of a thermoelectric material is quantified by the dimensionless figure of merit $ZT=S^2\sigma T/(\kappa_{lat}+\kappa_{el})$, where $S$ is the Seebeck coefficient, $\sigma$ electric conductivity, $\kappa_{lat}$ the lattice thermal conductivity, $\kappa_{el}$ the electric thermal conductivity, and $T$ the absolute temperature\cite{Wang2025,Snyder2008}. Achieving high thermoelectric performance therefore requires maximizing the power factor, $S^2 \sigma$ while simultaneously minimizing the total thermal conductivity. 
Since most high-performance thermoelectric materials are heavily doped semiconductors\cite{Ming2025}, where phonon-mediated heat transport dominates over the electronic contribution, the lattice thermal conductivity $\kappa_{lat}$ typically represents the major component of the total thermal conductivity. As a result, accurately predicting $\kappa_{lat}$ is crucial for screening and optimizing thermoelectric materials\cite{Snyder2008,He2017}.

Computing lattice thermal conductivity from first-principles typically requires either constructing large supercells to extract harmonic and anharmonic interatomic force constants (IFCs), which capture both phonon dispersions and three-phonon or four-phonon scattering processes, and subsequently solving the phonon Boltzmann transport equation (BTE) or performing long ab-initio molecular dynamics (AIMD) simulations to capture heat transport over extended timescales\cite{Jong2024,Gu2021}. Both approaches are computationally demanding, and even when first-principles calculations are substituted with MLIPs, the computational cost remains prohibitive, rendering large-scale screening of the extensive structural space produced by generative models infeasible.

To address this challenge, we evaluate a diverse set of surrogate models for predicting lattice thermal conductivity directly from given candidate structures. These include end-to-end deep neural networks that take atomic structures as input and learn task-specific representations through their architectures\cite{CrabNet,CGCNN}, feature-engineered models that rely on hand-crafted physicochemical descriptors with strong inductive bias\cite{KAN}, and models that leverage pre-trained universal MLIPs embeddings, which are subsequently used as input features to train downstream  feedforward neural networks. 

In this study, we create a benchmark by developing 15 surrogate models using the Phonix dataset\cite{auto_kappa}, which contains 6,966 entries with anharmonic phonon properties derived from first-principles calculations. Owing to its broad chemical and structural coverage, this dataset enables surrogate models to learn relationships between general structures and thermal transport properties, allowing rapid identification of low lattice thermal conductivity candidates and robust generalization beyond materials traditionally studied for thermoelectric applications. 
To systematically assess model performance and generalization, we train and evaluate each model under three dataset splits designed to probe qualitatively different aspects of predictive capability. First, a random split serves as the standard baseline, where training and test materials are drawn from the same overall distribution, assessing general interpolation accuracy. Second, a space-group disjoint split ensures that all crystal symmetries (space groups) present in the test set are entirely absent from the training set; this reflects a realistic scenario in which models are deployed on structurally novel materials and directly tests the ability to extrapolate across unseen structural motifs. Third, an out-of-distribution (OOD) split based on $\kappa_{lat}$ trains models exclusively on high-$\kappa_{lat}$ materials ($\kappa_{lat}$ $>$ 1 $\mathrm{Wm^{-1}K^{-1}}$) and evaluates them on low-$\kappa_{lat}$ materials ($\kappa_{lat}$ $\leq$ 1 $\mathrm{Wm^{-1}K^{-1}}$). This threshold is motivated by the thermal conductivity requirements of state-of-the-art thermoelectric materials, for instance, Bi$_{2}$Te$_{3}$ achieves zT $>$ 1 at room temperature in part due to its intrinsically low-$\kappa_{lat}$ of $\sim$1 $\mathrm{Wm^{-1}K^{-1}}$. This design simulates the realistic challenge of generalizing from the more abundant, high-conductivity materials in existing datasets to the rare, low-conductivity candidates most desirable for thermoelectric applications. Together, these three splits enable us to probe general predictive accuracy, structural extrapolation, and property-based extrapolation, respectively.
The best-performing models are identified as the most promising candidates for integrating with generative design workflows for accelerated thermoelectric materials discovery.
\begin{figure}[h]
    \centering
    \includegraphics[width=1.0\textwidth]{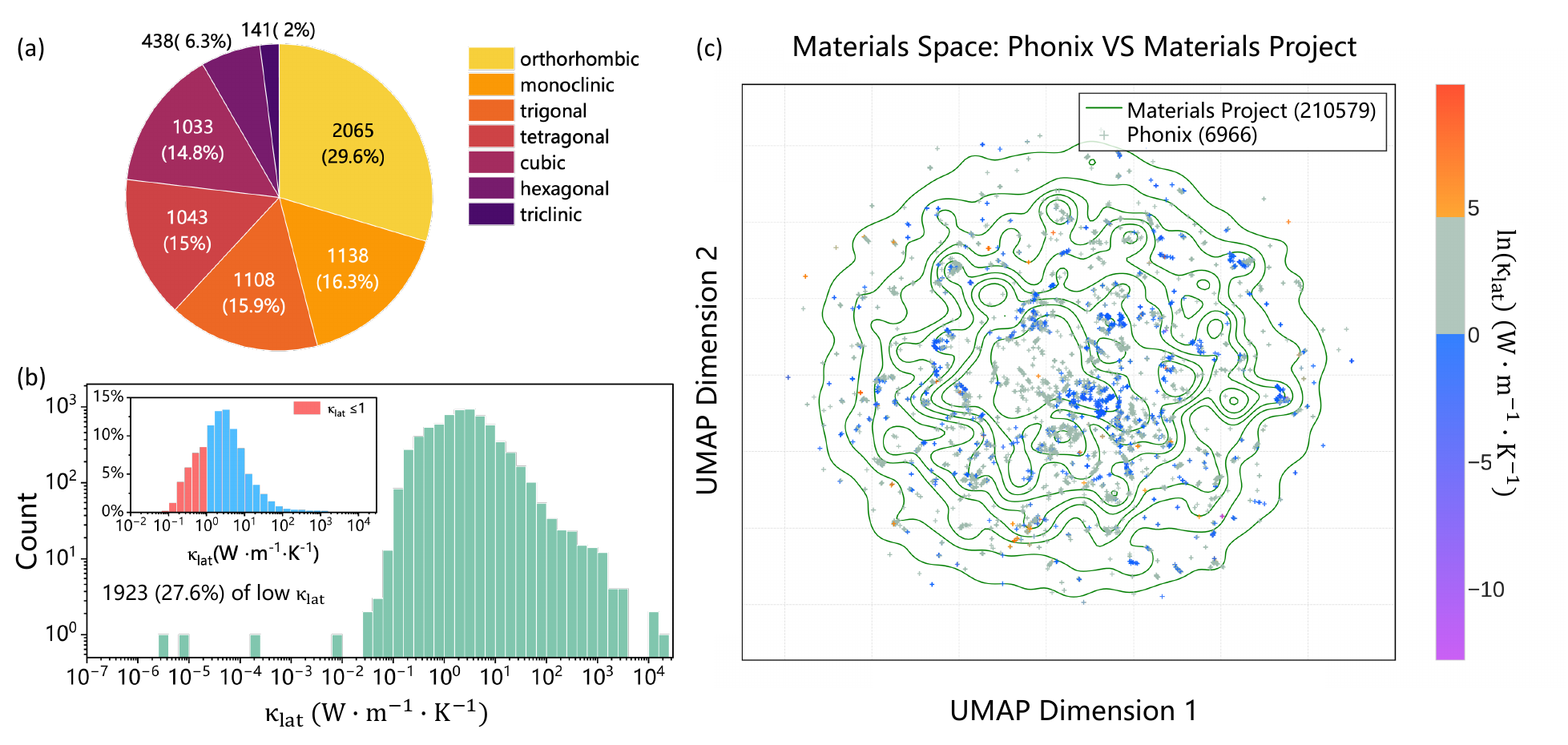}
    \caption{Material distribution in Phonix. (a) Distribution and percentage of crystal systems. (b) Distribution of $\kappa_{lat}$ at 300 K, derived from the Phonix database\cite{auto_kappa}. The subgraph represents the percentage of materials with different $\kappa_{lat}$. Low $\kappa_{lat}$ means that thermal conductivity values are lower than 1 $\mathrm{Wm^{-1}K^{-1}}$. (c) Umap for comparing with all Materials Project structures\cite{MP_database}.}
    \label{fig:distribution}
\end{figure}
\section{Methods}
\subsection{Lattice Thermal Conductivity}
In Phonix database, the lattice thermal conductivity is automatically evaluated through auto-kappa calculation framework\cite{Ohnishi2026}, which leverages the VASP (versions 6.4.2)\cite{vasp} and ALAMODE (versions 1.5)\cite{alamode} package where particle-like phonon transport is treated via the Peierls–Boltzmann transport equation and wave-like phonon coherence is captured using the Wigner transport equation\cite{HH_TC_DFT,DFT_4_Si}. All calculations follow an unified, automated high-throughput workflow, ensuring consistency and reproducibility across the entire dataset. \cref{fig:distribution}a shows the percentage of crystal systems in the Phonix materials, demonstrating broad crystallographic diversity and strong potential for model generalization across different symmetry classes. The statistical distribution of lattice thermal conductivity at \SI{300}{K} is presented in \cref{fig:distribution}b. The subgraph demonstrates that most materials (approximately 95\%) fall within the range of 0.14–39 $\mathrm{Wm^{-1}K^{-1}}$, while the dataset also includes a significant subset of low-thermal-conductivity compounds ($\kappa_{lat} <$ 1.0 $\mathrm{Wm^{-1}K^{-1}}$), totaling 1,923 entries. This distribution ensures sufficient representation of ultra-low-$\kappa_{lat}$ materials, which are critical for thermoelectric discovery.

\subsection{Dataset Split}
The dataset used for model training and evaluation is derived from the Phonix dataset\cite{auto_kappa}, which contains lattice thermal conductivity values for 6,966 inorganic crystalline materials. 

In thw Phonix dataset\cite{Ohnishi2026}, all structures originate from the PhononDB\cite{Phonondb} and the Materials Project\cite{MP_database} repositories. For each entry, lattice thermal conductivity was automatically computed using a standardized high-throughput workflow based on first-principles calculations and phonon transport simulations, providing temperature-dependent $\kappa_{lat}$ values.

\begin{figure}[h!]
    \centering
    \includegraphics[width=1.0\textwidth]{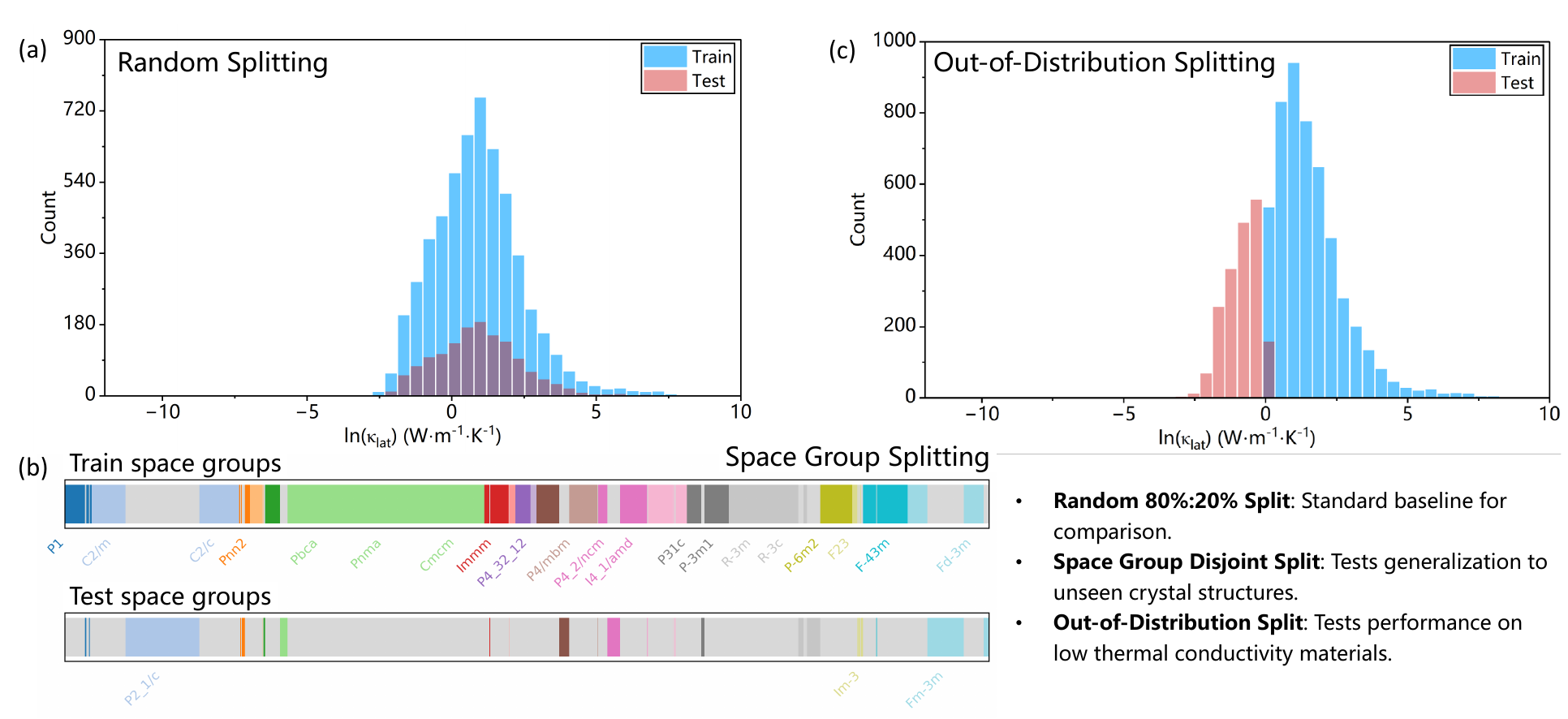}
    \caption{Dataset split for the benchmark. (a) Random 80\%:20\% split. (b) Space-group disjoint split. (c) Out-of-distribution split.}
    \label{fig:split}
\end{figure}
To illustrate the chemical and structural diversity covered by the Phonix dataset, we project the high-dimensional composition and structure-based features into a two-dimensional UMAP manifold\cite{umap}, as shown in \cref{fig:distribution}c. The features for each material were generated by Matminer package\cite{matminer}, including composition and structure features. Scatter points represent Phonix samples, while contour lines indicate the probability density of inorganic crystals in the Materials Project, estimated via kernel density estimation (KDE). Regions enclosed by inner contours are more densely populated in chemical space, indicating structurally similar and more commonly occurring crystal types. Note that the UMAP axes correspond to abstract nonlinear combinations of structural features. Phonix entries are broadly distributed across the manifold and largely overlap with the dense central region of the Materials Project distribution, indicating good representativeness with respect to commonly occurring crystal prototypes. This suggests that the surrogate models are more likely to generalize well within this region. Meanwhile, several Phonix materials appear in sparse peripheral areas, representing structurally rare OOD regions. Low-$\kappa_{lat}$ (blue points) compounds populate the central region of the UMAP manifold, where structurally complex and chemically diverse materials cluster, indicating strong phonon scattering and reduced rigidity of lattice data. In contrast, high-$\kappa_{lat}$ materials (orange points) mainly locate near the periphery, corresponding to highly symmetric and strongly bonded crystal frameworks that sustain more efficient phonon transport.

As shown in \cref{fig:split}, to systematically assess surrogate-model generalization, we adopt three distinct dataset splitting strategies based on $\ln(\kappa_{lat})$. (i) A random 80\%:20\% split is used as the standard baseline for evaluating in-distribution predictive performance. (ii) A space-group disjoint split is introduced to assess structural extrapolation, in which no crystallographic symmetry groups overlap between the training and test sets. (iii) An OOD split is constructed by assigning low-thermal-conductivity materials exclusively to the test set, enforcing a distributional gap and enabling a dedicated evaluation of model robustness in predicting ultra-low $\kappa_{lat}$ where strong anharmonicity and structural complexity make accurate regression particularly challenging. These three splits jointly provide a comprehensive protocol for benchmarking model performance under increasingly difficult extrapolation scenarios.
\begin{figure}[b!]
    \centering
    \includegraphics[width=1.0\textwidth]{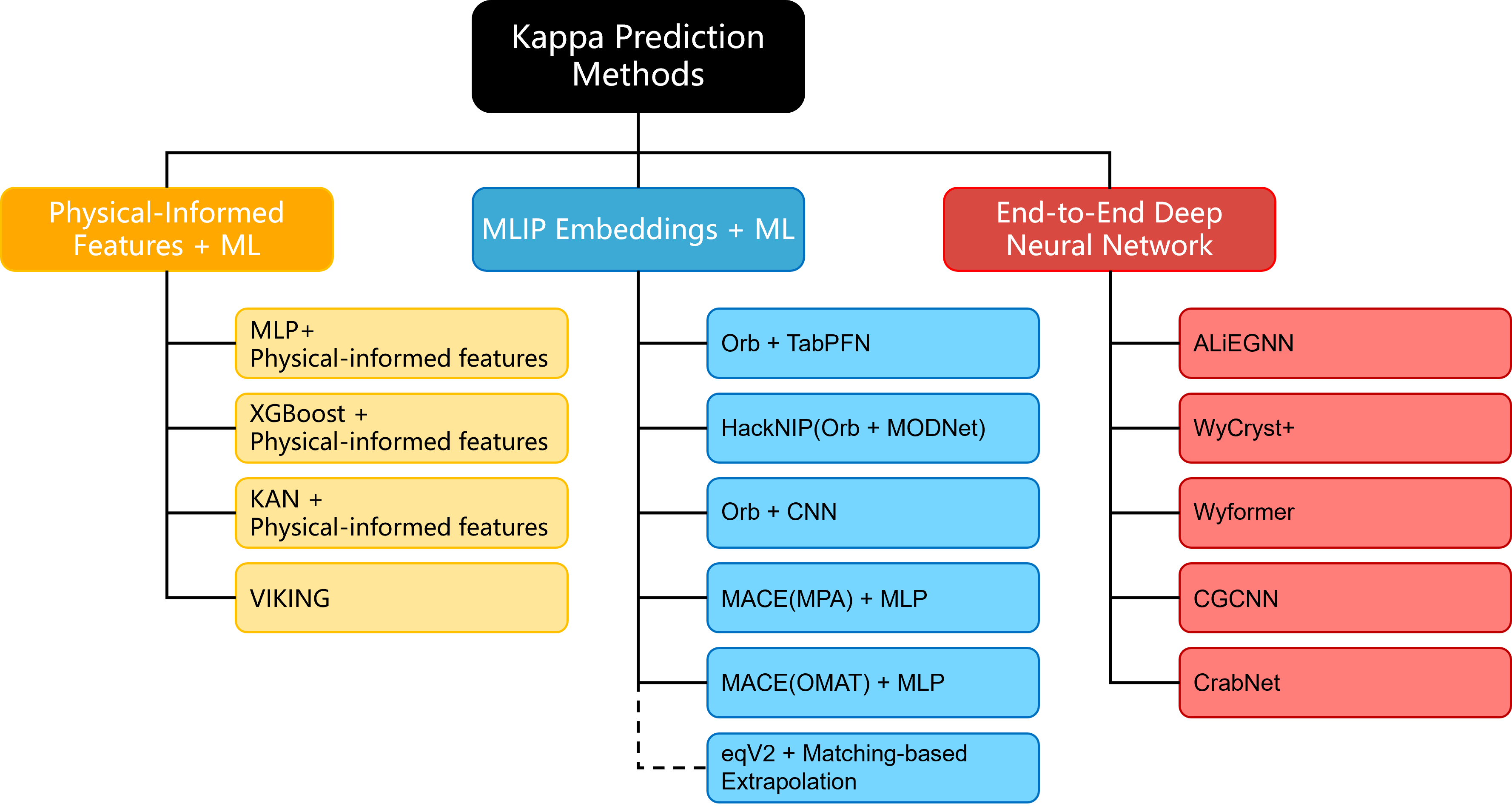}
    \caption{Overview of the surrogate models categorized into three groups, with the corresponding developers indicated for each model.}
    \label{fig:surrogate_overview}
\end{figure}
\subsection{Surrogate Models}
In this study, we develop and  evaluate 15 surrogate models for predicting lattice thermal conductivity under three distinct dataset splits. While several of the surrogate models are based on established architectures, the majority are modified or developed from existing frameworks to enhance predictive performance and generalization. These models span a range of input representations and learning strategies, enabling a systematic comparison of their effectiveness for lattice thermal conductivity prediction. For detailed descriptions of each surrogate model and their respective architectures, we refer the reader to the Supplementary Materials. 

An overview of the 15 surrogate models is presented in \cref{fig:surrogate_overview}. The surrogate models can be categorized into three distinct groups: 1) Physical-informed feature descriptors combined with ML models, 2) Deep neural networks based on structural representations, 3) MLIP-derived embeddings combined with ML models. 

Models in the first category employ chemically and/or physically motivated descriptors of the inorganic crystal structures and in some cases are augmented with structural information. These features are then used as inputs to machine learning regression models to predict lattice thermal conductivity. Models in the second category utilize structural and stoichiometric descriptions of inorganic crystal structures as inputs to deep neural networks, similar to those commonly employed in generative models and MLIPs. Models in the third category leverage pre-trained universal MLIPs\cite{batatia2023foundationmace,neumann2024orbfastscalableneural,rhodes2025orbv3atomisticsimulationscale} to extract learned representations of inorganic crystal structures. While no structural optimization is performed, the resulting MLIP-based embeddings are combined with feedforward neural networks to predict lattice thermal conductivity.

\section{Results}
\begin{table*}[htb!]
\centering
    \caption{Overall mean absolute error (MAE) of the predicted lattice thermal conductivity for all 15 surrogate models across the three distinct dataset splits, with MAE computed on the natural logarithmic scale of the target values.}  
    \resizebox{\linewidth}{!}{ 
    \begin{tabular}{ccccc}
    \toprule
     & Random split (MAE↓) & Space group split (MAE↓)& OOD split (MAE↓) &Avg MAE↓\\
    \midrule
    ALiEGNN& 0.379&	0.512&	1.245&	0.712\\
    Orb + CNN&	0.440&	0.510&	1.404&	0.784\\
    HackNIP\cite{Kim2025}&	0.380&	0.501&	1.516&	0.799\\
    ViKING& 0.487& 0.597& 1.370& 0.818\\
    CGCNN\cite{CGCNN}&	0.523&	0.652&	1.294&	0.823\\
    KAN + Physical-informed feat.&	0.473&	0.596&	1.424&	0.831\\
    MLP + Physical-informed feat.&	0.431&	0.627&	1.454&	0.837\\
    MACE(OMAT)+MLP&	0.448&	0.601&	1.466&	0.838\\
    MACE(MPA)+MLP&	0.445&	0.602&	1.484&	0.843\\
    WyFormer\cite{kazeev2024wyckofftransformer}&	0.503&	0.625&	1.419&	0.849\\
    XGBoost + Physical-informed feat.&	0.444&	0.540&	1.610&	0.865\\
    Crabnet\cite{CrabNet}&	0.523&	0.674&	1.422&	0.873\\
    Orb + TabPFN&	0.378&	0.494&	1.764&	0.879\\
    eqV2-MEX&	-&	-&	0.997&	-\\
    WyCryst+&	0.491&	0.700&	1.516&	0.902\\
        \bottomrule
    \end{tabular}
    }
    \label{table:results}
\end{table*}
By training each surrogate model on the three dataset splits, the predictive error for lattice thermal conductivity was obtained. The Mean Absolute Error (MAE) for all 15 surrogate models across the three dataset splits is reported in \cref{table:results}. Overall, ALiEGNN achieves the best performance among the evaluated models, while Orb+CNN and HackNIP closely follow, exhibiting reduced predictive accuracy primarily in the OOD dataset split.

In contrast WyCryst+ models show higher predictive errors across all dataset splits. Nevertheless, most surrogate models demonstrate robust performance across the different split strategies. Based on a comparative analysis of overall performance, a clear top-performing models can be identified.

\begin{figure}
    \centering
    \includegraphics[width=0.9\textwidth]{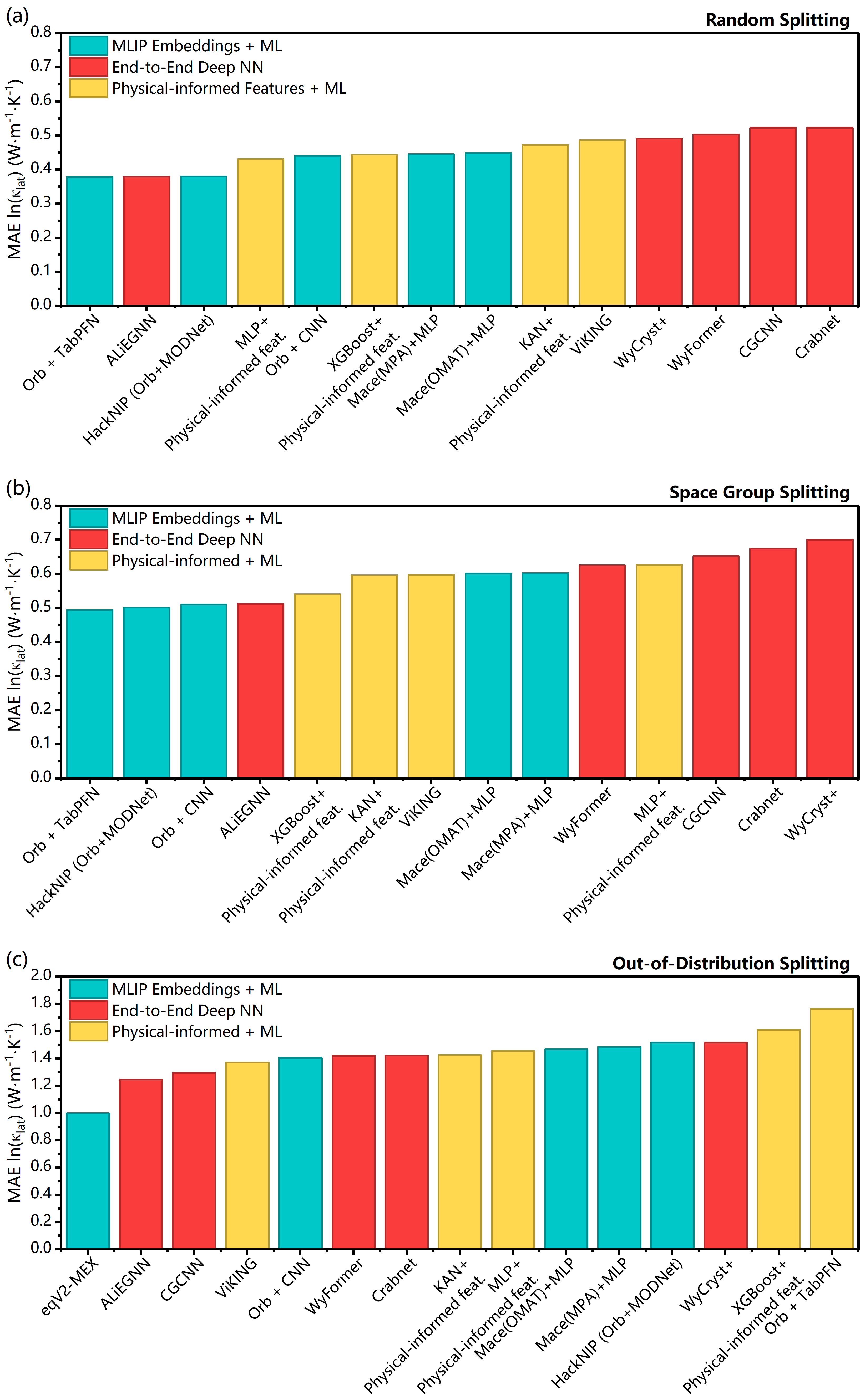}
    \caption{The performance across the 15 surrogate models for each individual data split, with color separation into the three identified categories.}
    \label{fig:results}
\end{figure}

In \cref{fig:results}, the predictive performance of each surrogate model is compared across the individual dataset splits, with models color-coded according to their respective categories. This representation enables assessment of both the performance of individual surrogate models within each dataset split and the overall behavior of each model category across different data regimes.

For the random split, MLIP-embedded models generally outperform the other approaches, with custom feature models closely following. Interestingly, deep neural network models exhibit the weakest performance in this split, with the notable exception of ALiEGNN, which ranks among the best-performing models. This likely stems from its explicit encoding of bond-angle information via spherical harmonics, allowing it to distinguish local environments that distance-only GNNs cannot. And this is an advantage well-suited to $\kappa_{lat}$ prediction, which depends sensitively on bond-angle-driven phonon dispersion and anharmonic scattering and also underlies the strong performance of MACE and other equivariant MLIPs in this benchmark.

A similar trend is observed for the space-group split, where MLIP-embedded models again achieve superior performance, while deep neural networks perform poorly overall, except for ALiEGNN. In contrast, for the OOD split, deep neural network models show the strongest performance, outperforming both custom feature models and MLIP-embedded models. In this regime, MLIP-embedded approaches rank among the weakest-performing models. This finding is consistent with Zhang et al\cite{ood_zhang}, who demonstrated that fine-tuning pretrained atomistic models can induce representation collapse. It's a degradation of chemically and geometrically meaningful priors, which may lead to impair OOD generalization. A complementary perspective is offered by Omee et al\cite{Omee2024}, whose large-scale OOD benchmark for materials property prediction found that simpler graph neural networks such as CGCNN and ALIGNN exhibit more robust generalization than more complex architectures. This aligns well with our observations, where CGCNN and the ALIGNN-derived ALiEGNN emerge as strong performers on the OOD thermal conductivity dataset.

This indicates that MLIP-embedded models excel at interpolation within well-sampled chemical and structural spaces but struggle to generalize to strongly OOD regimes. In contrast, deep neural network models based on explicit structural representations demonstrate superior extrapolation capability, making them more suitable for predicting lattice thermal conductivity in unexplored regions of materials space.

Notably, when comparing three distinct deep neural network baseline models, CGCNN (structure baseline), WyFormer (Wyckoff baseline), and CrabNet (composition baseline), we observe a systematic degradation in lattice thermal conductivity prediction performance that directly correlates with decreasing representation  expressivity. Moving from full structural information to Wyckoff-level symmetry and finally to composition alone, the MAE increases consistently across all data splits. This trend aligns with physical intuition: lattice thermal conductivity is primarily governed by detailed crystal structure, and thus models with richer structural representations naturally outperform those with more abstract or reduced feature sets.

The custom feature models show consistent and competitive performance across all dataset splits, indicating that chemically and physically motivated descriptors provide robust predictive signals for lattice thermal conductivity especially in small data regime, although they are generally outperformed by the other models.

The key advantage of surrogate models lies in their significantly lower overall computational time compared to DFT method. As the most accurate method, DFT calculation consumes substantial time in solving force constants and other processes. 
Our ALiEGNN required only 2,750 seconds for model training, while its inference time for the test set did not exceed 5 seconds, showing its substantially lower computational cost compared to DFT. In summary, employing surrogate models for high-throughput screening across vast material spaces represents the optimal choice for balancing prediction accuracy and computational cost.



\section{Conclusion}
In this work, we use the Phonix database \cite{auto_kappa}, comprising 6,966 inorganic compounds with computed anharmonic phonon properties, to evaluate the predictive accuracy of 15 surrogate models for lattice thermal conductivity. The models are assessed using three complementary dataset splits: i) a random split, ii) a space-group disjoint split, and iii) an Out-Of-Distribution(OOD) split. These splits enable evaluation of not only individual surrogate models but also their broader methodological classes, which are categorized into three groups: 1) Physical-informed feature descriptors combined with ML models, 2) End-to-end deep neural networks based on structural representations and 3) MLIP-derived embeddings combined with ML models. 

We find that ALiEGNN achieves the best overall predictive performance, while Orb+CNN and HackNIP closely follow, exhibiting higher MAE primarily in the OOD dataset split. MLIP-derived embedding models demonstrate excellent interpolation accuracy within well-sampled regions of materials space but perform poorly under OOD conditions. In contrast, end-to-end deep neural network models show weaker interpolation performance but superior extrapolation capability. Motivated by these complementary strengths, we find that an ensemble combining models from these two categories yields improved and more robust performance across all dataset splits.

Although none of the surrogate or ensemble models come close to the accuracy of direct DFT-based lattice thermal conductivity calculations across all datasets, their substantially lower computational cost provides a decisive advantage. The speed–accuracy trade-off offered by surrogate models enables efficient high-throughput screening of thermoelectric materials, with only minimal loss in predictive accuracy.

\textbf{Acknowledgements}
K.H. acknowledges funding from the MAT-GDT Program at A*STAR via the AME Programmatic Fund by the Agency for Science, Technology and Research under Grant No. M24N4b0034. This work was partly supported by JST SPRING, Grant Number JPMJSP2108. This research was partly supported by the National Research Foundation, Prime Minister’s Office, Singapore under its Campus for Research Excellence and Technological Enterprise (CREATE) programme. Numerical calculations were performed using the following supercomputers through the HPCI System Research Project (Project IDs: hp240194): Grand Chariot at the Information Initiative Center, Hokkaido University; SQUID at the D3 Center, Osaka University; Wisteria/BDEC-01 at the Supercomputing Division, Information Technology Center, The University of Tokyo. Additional resources were provided by the Supercomputer Center, Institute for Solid State Physics, The University of Tokyo, and MASAMUNE-IMR at the Center for Computational Materials Science, Institute for Materials Research, Tohoku University.

\textbf{Data and code accessibility}
The dataset used for machine learning prediction is available in the GitHub repository at \url{https://github.com/masato1122/phonon_e3nn}. The Phonix database for anharmonic phonon interactions is available on ARIM-mdx at \url{https://phonix-db.org}, and code based used for this work, will be public upon release. Early access can be permitted by writing to the authors.

\bibliography{refs}
\newpage

\begin{appendices}

\section{Surrogate Models}
\subsection{Physical-informed features + MLP, XGBoost, KAN}
\subsubsection{Feature Engineering}

All three models utilize identical feature representations to enable direct performance comparison. The feature engineering pipeline incorporates three complementary descriptor categories that capture compositional, structural, and symmetry information. The first category comprises composition-based features extracted using the Matminer computational materials science toolkit. These features encode elemental properties and stoichiometric relationships, providing element-based compositional descriptors that capture atomic-level characteristics relevant to thermal transport. The second category consists of structure-based features derived from Smooth Overlap of Atomic Positions (SOAP) descriptors. SOAP features provide rotationally and permutationally invariant representations of local atomic environments, capturing three-dimensional structural information through smooth density distributions. These descriptors enable the models to learn structure-specific fingerprints that encode geometric arrangements of atoms. The third category encompasses symmetry-based features utilizing the Binary matrices representation. These features encode space group information through symmetric matrix representations, alongside Wyckoff position descriptors, space group symbols, and space group numbers.

All features undergo standardization to ensure zero mean and unit variance, which improves numerical stability and optimization convergence for gradient-based methods.

\subsubsection{XGBoost}

The XGBoost regressor represents an ensemble learning approach based on gradient-boosted decision trees. It employs mean squared error (MSE) as the loss function, optimized directly on the log-transformed target values. Hyperparameter optimization was performed through exhaustive grid search with cross-validation to identify optimal configurations. Model evaluation was conducted using mean absolute error (MAE) and coefficient of determination (R²), both computed on the log scale.

\subsubsection{MLP}

The Multi-Layer Perceptron (MLP) implements a conventional fully-connected feedforward neural network architecture. It constructs a deep learning regression model through stacked layers of linear transformations followed by nonlinear activation functions. The model comprises an input layer with dimensionality matching the featurized input space, multiple hidden layers implementing fully-connected transformations with nonlinear activations (typically Rectified Linear Units or similar functions), and a single-neuron output layer for regression. Training employed mean squared error as the loss function, optimized using either Adam or Stochastic Gradient Descent with momentum.

\subsubsection{KAN}

Kolmogorov-Arnold Network (KAN)\cite{KAN-origin} represent a novel neural network architecture that replaces traditional linear layers with learnable activation functions on network edges. This architecture is motivated by the Kolmogorov-Arnold representation theorem, which establishes that any multivariate continuous function can be represented as a superposition of continuous functions of a single variable. This design enables the network to learn optimal nonlinear transformations specific to each connection, potentially providing greater representational flexibility with fewer parameters compared to traditional architectures. The model accepts input tensors matching the dimensionality of the featurized dataset and produces single-valued outputs corresponding to log klat predictions. Training employed mean squared error as the loss function with gradient-based optimization, typically using the Adam optimizer.

\subsection{VIKING}
VIKING is a Bayesian graph neural network (BGNN) developed for the direct prediction of lattice thermal conductivity from crystal structures. The model is designed to balance structural expressivity with calibrated extrapolation behavior, particularly under distributional shifts such as space-group–disjoint and OOD dataset splits.

The architectural design of VIKING is conceptually inspired by prior advances in graph-based deep learning for molecular and materials modeling, particularly the development of graph neural networks for quantum chemistry and atomistic property prediction \cite{Glimer2017}, as well as subsequent symmetry-aware deep learning approaches for scientific machine learning \cite{Lim2021}. These studies established the importance of structured graph representations and symmetry-consistent learning for accurately modeling materials properties.

In addition, VIKING emphasizes ensemble Bayesian neural network architectures, structured crystallographic feature integration, and systematic evaluation across random, space-group, and OOD dataset splits. In particular, it incorporates Wyckoff-aware features, symmetry descriptors, dropout-based uncertainty regularization, batch normalization, and cross-validation–based model selection. 

Each crystal structure is represented as a periodic atomic graph, where atoms correspond to nodes and neighboring atoms within a cutoff radius define edges. Node features consist of learned elemental embeddings derived from atomic-number–dependent properties. Edge features encode interatomic distances expanded using radial basis functions (RBFs), enabling smooth distance-sensitive interactions while preserving translational invariance.

Message passing proceeds through stacked residual graph convolution layers:

\begin{equation}
h_i^{(l+1)} = h_i^{(l)} + \sum_{j \in \mathcal{N}(i)}
\phi_e(e_{ij}) \odot \phi_h\left(h_j^{(l)}\right)
\end{equation}

where $\phi_e$ and $\phi_h$ are multilayer perceptrons applied to edge features $e_{ij}$ and neighboring node features $h_j^{(l)}$, and $\odot$ denotes element-wise modulation. Residual connections are employed to stabilize training and preserve low-level structural information.

After $L$ message-passing layers, atomic embeddings are aggregated via normalized summation pooling to produce a global crystal representation:

\begin{equation}
z = \mathrm{Pool}\left({h_i^{(L)}}\right)
\end{equation}

This pooled representation is passed to a Bayesian regression head that predicts $\ln(\kappa_{\mathrm{lat}})$.

Unlike deterministic graph neural networks, VIKING models the final regression layers within a Bayesian framework. Instead of learning point estimates of the weights, approximate posterior distributions over weights are inferred using variational inference:
\begin{equation}
p(w \mid \mathcal{D}) \approx q_{\theta}(w)
\end{equation}
Training optimizes the evidence lower bound (ELBO):
\begin{equation}
\mathcal{L}_{\mathrm{ELBO}} =
\mathbb{E}_{q(w)} \left[ \log p(\mathcal{D} \mid w) \right] -\mathrm{KL}\left(q(w)||p(w)\right)
\end{equation}

This formulation introduces implicit regularization and calibrated predictive uncertainty. In sparsely sampled regions of structure or property space, posterior variance increases, reducing the risk of overconfident extrapolation. This behavior is particularly relevant for low $\kappa_{\mathrm{lat}}$ materials, which often occupy distributionally sparse regions characterized by strong anharmonicity and structural complexity.

VIKING was developed using an iterative rapid-prototyping workflow that combined domain expertise in materials graph learning with AI-assisted code generation and refinement. This collaborative development approach enabled rapid architectural experimentation, debugging of Bayesian training dynamics, and efficient hyperparameter optimization while maintaining a foundation in established graph neural network and variational inference theory.

\subsection{ALiEGNN}
\begin{figure}
    \centering
    \includegraphics[width=0.7\textwidth]{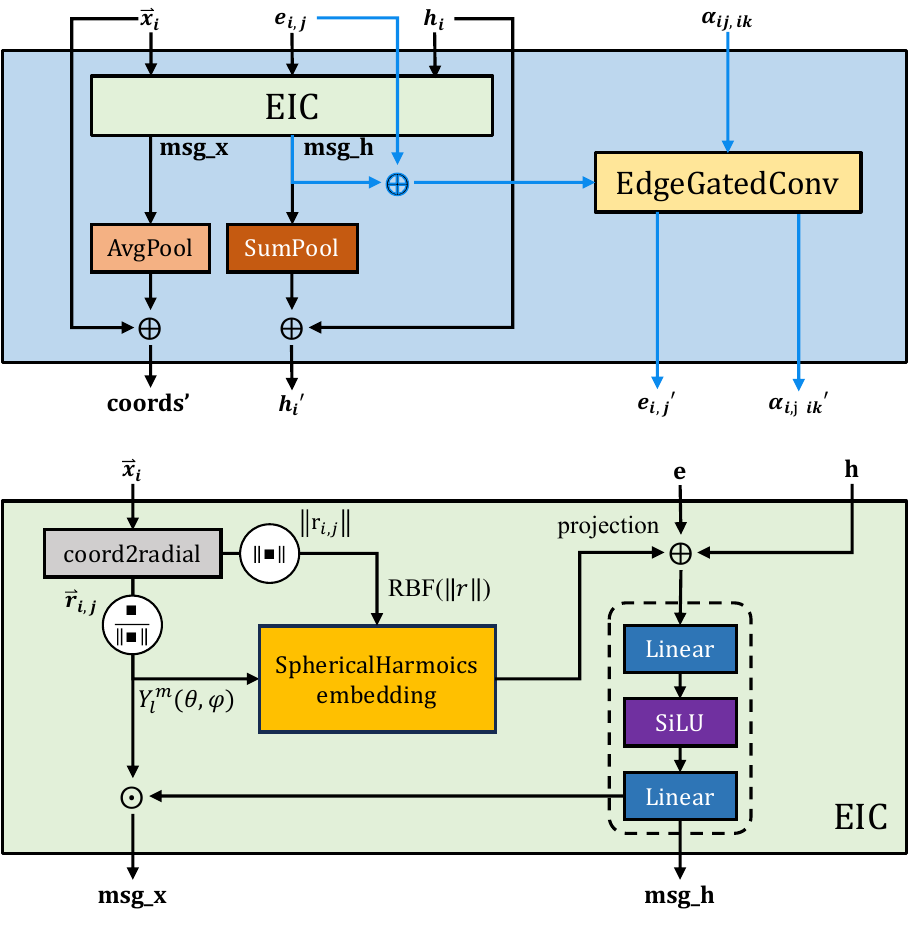}
    \caption{Schematic of the information passing layer in ALiEGNN, which jointly updates atomic coordinates, atom-wise features, bond and bond angle representations by integrating radial and angular embeddings with edge-gated message passing. Three layers are stacked in the ALiEGNN model.}
    \label{fig:ALiEGNN}
\end{figure}

Atomistic Line Equivariant Graph Neural Network (ALiEGNN) is an E(3)-equivariant graph neural network that integrates the message-passing framework of EGNN with angle information interactions inspired by ALiGNN through a novel Equivariant Interaction Convolution (EIC). The model is designed to enhance directional sensitivity in interatomic message passing while preserving strict Euclidean equivariance, making it suitable for modeling phonon-related properties in crystalline materials. A schematic overview of the architecture is shown in \cref{fig:ALiEGNN}.
For each atomic pair, the relative position vector ($\vec{r}_{i,j}$) is decomposed into its radial and angular components. The radial dependence is encoded using a set of radial basis functions applied to the interatomic distance ($RBF(\|r\|)$). The angular dependence is captured via real spherical harmonics ($Y_{l}^{m} (\hat{r}_{i,j})$), providing a rotation-equivariant representation of bond orientations.
The radial and angular embeddings are combined through a learned projection module consisting of linear layers and SiLU nonlinearities. The resulting edge features are then injected into the message-passing process via an edge-gated convolution, following the philosophy of the ALiGNN framework. In this formulation, bond angle information is propagated to update bond (edge) representations, which in turn modulate the information exchanged between atoms.
Compared to standard EGNN, ALiEGNN explicitly incorporates angular information via spherical harmonics, enabling the model to distinguish local environments that are indistinguishable under distance-only representations. As a result, ALiEGNN combines the strengths of both frameworks: the stability and simplicity of EGNN and the expressive angular sensitivity of ALiGNN.

\subsection{WyCryst+}
WyCryst+ is a Wyckoff-based, symmetry-aware crystal structure representation designed for surrogate property prediction, where a crystal is encoded as an ensemble of Wyckoff sites within the unit cell. In this representation, each Wyckoff position is described with explicit site-resolved information, enabling a structured input that is well suited for convolutional neural network learning. This Wyckoff position informed design was first introduced in Dis-GEN\cite{dis_gen} for disordered crystal generation, where it demonstrated robust capability in describing both ordered and disordered crystal structures. Based on this representation, WyCryst+ adapts the original WyCryst architecture \cite{zhu2024wycryst} by retaining only the forward prediction branch. The Wyckoff site input is processed through a CNN encoder to extract hierarchical structural features, and the resulting latent representation is then passed into a multilayer perceptron to perform forward prediction of target material properties. By combining a detailed Wyckoff-site representation with a CNN + MLP predictor, WyCryst+ provides an efficient and symmetry-consistent surrogate model for learning structure property relationships in both ordered and disordered crystals.

\subsection{Wyformer}
WyFormer is a Transformer-based model that represents crystal structures using a discrete, symmetry-aware Wyckoff sequence consisting of space group, element types, and Wyckoff site information. This representation avoids explicit atomic coordinates and encodes crystal structure at the level of symmetry and site occupancy. For property prediction, the input Wyckoff tokens are first embedded and processed by the Transformer to produce contextualized token representations. The resulting token embeddings corresponding to atomic sites are then aggregated into a single crystal-level representation using a weighted pooling operation, where the weights reflect the multiplicities of the associated Wyckoff positions. This pooled representation is passed through a lightweight multilayer perceptron to predict scalar material properties. By operating directly on Wyckoff-based structural tokens, WyFormer enables property prediction from symmetry and composition alone, without requiring full geometric information.

\subsection{CGCNN}
CGCNN represents a periodic crystal as an undirected graph, where atoms are nodes and interatomic connections are edges. Each atom is described by a fixed atomic feature vector encoding elemental properties, while each bond is characterized by a feature vector that includes discretized interatomic distance information. The model applies a series of graph convolution layers in which atomic features are iteratively updated by aggregating information from neighboring atoms and their connecting bonds. Neighbor contributions are modulated by a learnable gating mechanism that determines the importance of each interatomic interaction based on both atomic and bond features. This gated message passing allows CGCNN to adaptively weight different local coordination environments. After several convolution layers, the learned atomic representations are pooled using a permutation-invariant normalized summation to obtain a global crystal representation. This representation is then passed through fully connected layers to predict target material properties. 

\subsection{CrabNet}

CrabNet (Compositionally Restricted Attention-Based Network) \cite{CrabNet} is a structure-agnostic model that predicts materials properties from chemical composition alone (chemical formula), serving in this benchmark as the composition baseline. It uses a transformer-style self-attention mechanism to learn context-dependent element representations and their contributions to the target property.

Each composition is featurized as an element-derived matrix (EDM): rows correspond to elements and columns to embedding dimensions. Element identities are encoded using learned embeddings (default: mat2vec); stoichiometric information is encoded via fractional embeddings (linear and log-scale) so that small concentrations (e.g., dopants) are preserved. The EDM is passed through a transformer encoder with three layers and four attention heads per layer. Each head performs scaled dot-product self-attention (query, key, value projections), updating element representations based on inter-element interactions. The final element representations are transformed by a fully connected residual network into per-element contributions; the predicted property is the mean of these contributions. The same residual pathway also yields aleatoric uncertainty estimates.

Training uses the robust MAE loss, the look-ahead and Lamb optimizers, and a learning rate cycled between $1\times10^{-4}$ and $6\times10^{-3}$ every 4 epochs. Batch size is chosen dynamically within 27--212 based on dataset size. For reproducibility, a fixed random seed (42) is used. Target values are standardized to zero mean and unit variance during training and unscaled for evaluation. In this work, CrabNet is applied to predict $\ln(\kappa_{\mathrm{lat}})$ from composition only, without structural or Wyckoff information, to assess the limit of composition-only surrogates for lattice thermal conductivity.

\subsection{HackNIP}
HackNIP \cite{Kim2025} adopts a two-stage architecture that integrates (i) feature extraction using a pretrained machine-learning interatomic potential (MLIP) and (ii) a lightweight downstream regressor for structure-to-property prediction. Specifically, an input atomic structure is first encoded into a fixed-length embedding by extracting node-level atomic features from intermediate layers of the pretrained MLIP. These embeddings are then aggregated and passed to a conventional regression model to predict the target property.

This hybrid design is intended to leverage the rich, transferable representations learned during large-scale MLIP pretraining while retaining the data efficiency and rapid training afforded by shallow regression models. In the original work, multiple combinations of MLIPs (Orb-v2, Equiformer, and MACE) and downstream regressors (MODNet, XGBoost, and a multilayer perceptron) were systematically evaluated. Based on predictive performance, the Orb-v2–MODNet combination was identified as the optimal configuration and is therefore adopted in this work for lattice thermal conductivity prediction.

We followed the preprocessing and embedding extraction as demonstrated in the original work, where node embeddings of Orb-v2 across layers raging L = 1-15 were extracted. This produces a hierarchical family of embeddings: shallow layers primarily capture local atomic environments, whereas deeper layers increasingly encode more global structural context. For each layer depth, Orb-v2 node embeddings were mean-pooled to produce fixed-length 256-dimensional embeddings, after which the optimal depth was determined based on validation performance. 

To determine the best layer depth for our downstream task, we trained a MODNet regressor with four fully connected layers (256 neurons each) separately on the embeddings from each depth using the random split. Among the evaluated depths, layer L = 11 yielded the best performance. We further optimized the downstream MODNet regressor based on validation performance and report the configuration that minimizes the cross-validated MAE. In particular, we perform the Optuna (TPE) hyperparameter optimization using the selected embedding depth L = 11 (see Table 2). The same setting were used to get the results reported for space-group split and the OOD split. 

\begin{table*}[htb!]
\centering
\small
\renewcommand{\arraystretch}{1.2}

\begin{tabular}{lccccccc}
\toprule

\multirow{2}{*}{Split} &
\multirow{2}{*}{\makecell{Orb-v2\\ depth $L^*$}} &
\multicolumn{3}{c}{MODNet model parameters} &
\multicolumn{3}{c}{MODNet fitting parameters} \\

\cmidrule(lr){3-5} \cmidrule(lr){6-8}

& & $n_{\text{feat}}$ & num\_neurons & out\_act &
batch\_size & lr & loss \\

\midrule

Random Split& 11 & 247 & ([137],[137],[ ],[ ]) & linear & 64 & 0.001195 & mae \\

\bottomrule
\end{tabular}

\caption{Optimal Orb-v2 layer embedding depth and MODNet hyperparameters.}
\end{table*}

\subsection{Orb+CNN}
In this work, ``Orb+CNN'' refers to the sequence-based CNN that operates on per-atom ORB embeddings with masked global pooling. Crystal structures are converted to atom graphs and passed through the ORB model \cite{rhodes2025orbv3atomisticsimulationscale}; the encoder used is \texttt{orb-v3-conservative-20-omat}, which outputs 256-dimensional node features per atom (no pooling at the encoder stage).

The downstream model is a one-dimensional convolutional network applied to variable-length sequences of these 256-dimensional vectors. The input is arranged as (batch, 256, $n_{\text{atoms}}$). The convolutional stack consists of three layers: 256$\to$128 channels (kernel size 3, padding 1), 128$\to$64 (kernel 3, padding 1), and 64$\to$32 (kernel 2, padding 0), each followed by LeakyReLU (negative slope 0.1) and dropout (0.2). Masked global average pooling over the atom dimension yields a 32-dimensional vector, which is passed through fully connected layers 32$\to$64$\to$32$\to$16$\to$1. Optionally, five material-property features (formation energy, energy above hull, band gap, bulk and shear moduli) can be concatenated after the first FC layer. The output is the predicted $\ln(\kappa_{\mathrm{lat}})$.

Training uses the Adam optimizer with learning rate $8\times10^{-5}$, batch size 32, and up to 100 epochs with early stopping (patience 15 on validation loss). The loss is mean squared error on $\ln(\kappa_{\mathrm{lat}})$, with an optional weighted log-MAE variant that emphasizes low-$\kappa_{\mathrm{lat}}$ materials.

\subsection{Orb+TabPFN }
Orb+TabPFN combines the same ORB encoder as Orb+CNN (\texttt{orb-v3-conservative-20-omat} \cite{rhodes2025orbv3atomisticsimulationscale}) with the TabPFN tabular regressor \cite{hollmann2025accurate}. Per-atom 256-dimensional node features are extracted for each crystal; no layer selection is applied. For each structure, masked statistics over the atom dimension are computed: mean, standard deviation, and maximum over the 256 dimensions, yielding a 768-dimensional feature vector (256$\times$3). Optionally, material-property features can be concatenated. If the total number of features exceeds 60, principal component analysis (PCA) is applied to reduce dimensionality to 60 components before feeding the data to TabPFN.

The TabPFN regressor (\texttt{TabPFNRegressor} from the \texttt{tabpfn} package) is a prior-fitted tabular transformer that is fit once without an explicit epoch count; training runs on GPU when available, otherwise on CPU. TabPFN imposes limits on the number of features (the implementation uses PCA when features exceed 60) and on the number of training samples (e.g., 10,000 on GPU and 1,000 on CPU). When the training set exceeds the sample limit, a random subsample is drawn with a fixed random seed to ensure reproducibility.

\subsection{MACE+MLP}
Utilizing two foundation models based on the MACE architecture \cite{batatia2022mace,batatia2023foundationmace} (MACE v0.3.13), MACE-MPA-medium and MACE-OMAT-medium, the node features from the final layer of the network were extracted for all candidate thermoelectric materials. In the standard MACE framework, these node features are typically aggregated via a summation to predict the total energy, while forces and stresses are obtained through gradients with respect to atomic positions \cite{unke2021machine}.

\begin{figure}
    \centering
    \includegraphics[width=0.7\textwidth]{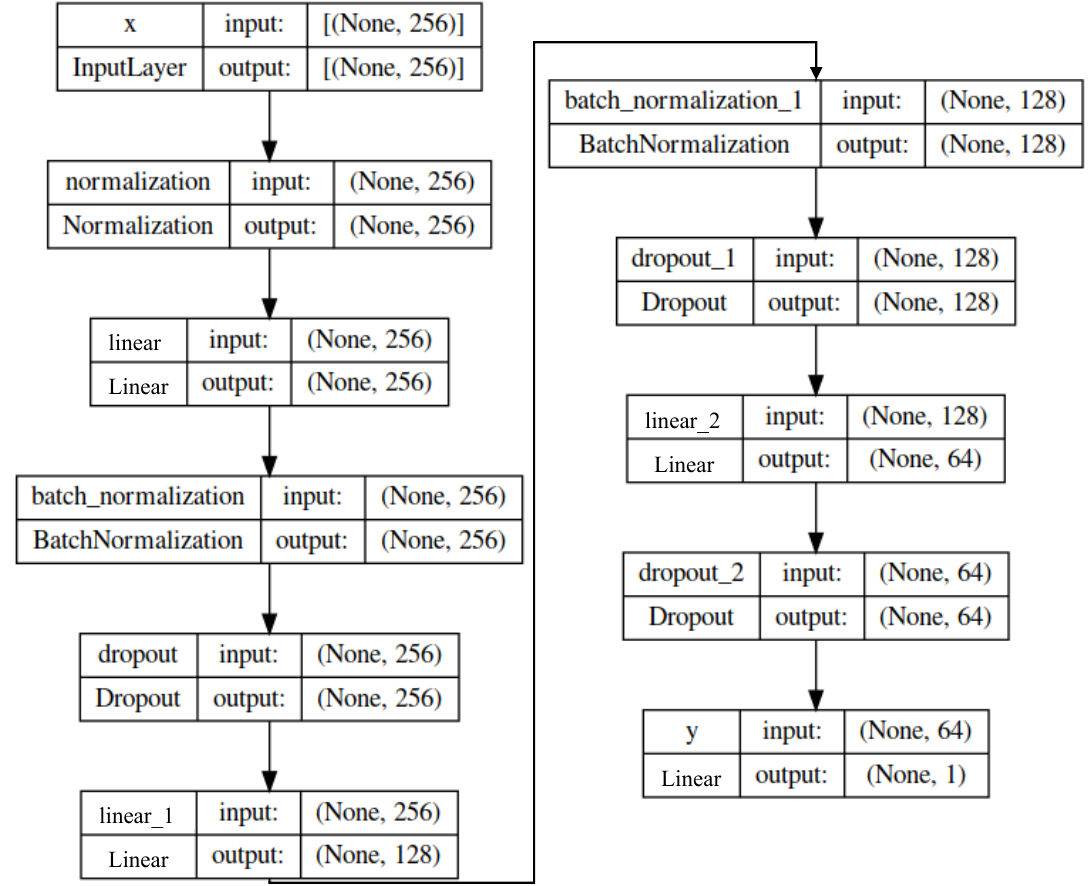}
    \caption{Schematic of the MLP architecture used to predict the lattice thermal conductivity from node features extracted from the MACE foundation models for the candidate materials. Batch normalization is applied after each linear layer, together with a dropout rate of 0.2, except for the final layer where a dropout rate of 0.1 is used. The "None" indicate the batch size, which is set to be 64.}
    \label{fig:mace_mlp}
\end{figure}

In this work, however, the node features were directly used as descriptors of the potential energy landscape for each candidate material. These descriptors encode information learned from the underlying training datasets of the foundation models, motivating the use of two models trained on distinct data distributions to capture complementary aspects of the energy landscape.

Based on the extracted node features, a Multilayer Perceptron (MLP) was trained to predict the lattice thermal conductivity. A schematic overview of the model architecture is shown in Fig. \cref{fig:mace_mlp}. The MLP was trained using the Adam optimizer \cite{kingma2014adam} with a learning rate of $5e-3$ and the mean absolute error (MAE) as the loss function. The same model architecture and training hyperparameters were used for node features extracted from both foundation models and across all three data splits, ensuring consistency and avoiding bias in the training procedure.

\subsection{eqV2 + Matching-based Extrapolation (MEX)}
In the eqV2 + MEX setup \cite{mianzhi2025on}, the EquiformerV2 (eqV2) backbone is used as the material feature encoder, and the Matching-based EXtrapolation (MEX) framework is applied on top for property prediction. In this approach, eqV2 first processes the input crystal structure to produce a fixed-dimensional material representation using its equivariant attention layers that respect geometric symmetries. Instead of predicting a scalar property directly through a conventional regression head, the MEX framework reframes property prediction as a matching problem between the encoded material representation and candidate property values. During training, MEX jointly learns: (1) the material encoder (here eqV2) that maps each crystal to a continuous latent vector, and (2) a label encoder that embeds property values into the same latent space. A similarity function (typically cosine similarity after optional projection) scores how well a material representation matches a property value. MEX is optimized using both absolute matching (pulling paired material and true label embeddings closer) and contrastive matching (distinguishing true labels from nearby noisy labels) to shape the shared latent space.  

At test time, given an encoded material from eqV2, the model performs property prediction by searching for the label value with the highest matching score in the learned latent space rather than via a direct regression output. This approach decouples the backbone representation from numeric regression and improves extrapolative generalization.  

\label{app:Training}

\end{appendices}

\end{document}